**D. S. Kovalenko\*, V. V. Kobychev, L. M. Kobycheva, O. V. Zueva**

*Institute for Nuclear Research, National Academy of Sciences of Ukraine, Kyiv, Ukraine*

# SIMOURG2.0 – GEANT4 APPLICATION FOR SIMULATION OF NUCLEAR DETECTORS WITH SIMPLE GEOMETRIES[1]

Simourg is a software based on the Geant4 toolkit and provides the Monte Carlo simulation of nuclear spectrometric setups with simple geometries for such applications as nuclear decay research, radiation safety, and nuclear medicine. With just a few command lines, users can define simple geometries, materials, and radiation sources to obtain reliable approximations for typical experimental setups. Simourg version 2.0, currently in its prerelease stage, introduces extended functionality for data extraction, geometry configuration, debugging, and visualization.

*Keywords*: Monte Carlo simulation, radiation detector, nuclear spectrometry, detector geometry, detector response, energy deposition, particle trajectory, 3D visualization.

## 1. Introduction

Nuclear spectrometry plays a crucial role in numerous fields, including exotic nuclear decay research, radiation safety on nuclear plants, isotopes production, and nuclear medicine, where precise measurement of spectra of the ionizing radiation is essential. These applications often require accurate determination of detector efficiency and response functions, which are typically obtained through direct measurements accompanied by Monte Carlo (MC) simulation. Moreover, to pre-evaluate the experiment outcome, it's a useful practice to simulate different variants of the experiment setup and choose which will fit the most. While existing tools for MC simulation, like Geant4 [1] and MCNP, provide comprehensive capabilities, they are often overly complex for the relatively simple geometries used in many spectrometric setups.

To support simplified and efficient modeling of nuclear spectrometric detectors, the Simourg (**Sim**ulator **o**f **U**sually **R**equested **G**eometries) software was developed [2], based on the Geant4 Monte Carlo simulation framework. Using a simple text-based configuration file, users can define the geometry, material properties, source characteristics, and output parameters to calculate the detector's response function, making it especially suitable for low-background experiments and

---

routine spectrometric tasks. Thus, Simourg can be considered as a shell for Geant4-based simulations, encapsulating implementation details, hiding them from the user, and allowing one to focus on detector design and on its physics rather than on programming. Its physics accuracy is ensured by the validated models provided by Geant4, making Simourg a practical tool for rapid and reliable detector response simulations.

## 2. Simourg description

There are several tools similar to Simourg, including GAMOS [3] for medical-physics applications and the simplified multipurpose Geant4 interfaces such as GGS [4] and GEARS [5]. Among these, Simourg is closer to GAMOS, as it relies on dedicated scripts to implement full simulations with the main focus on nuclear spectrometry. By contrast, GGS and GEARS use a mixed approach – integrating GDML, scripting, and C++ objects – to support general-purpose simulations across a broad range of research fields.

To effectively model a nuclear spectrometric detector and develop its response function using the Simourg software, a structured approach is applied, encompassing geometry construction, source configuration, and simulation execution.

The geometry in Simourg is defined through a hierarchical list of objects that represent the detector setup. Each object in the configuration file corresponds to a geometric element – such as a cylinder or box – with specified dimensions, axial position, and material properties (composition and density). These elements can be nested, enabling users to construct complex assemblies like an active scintillation crystal placed inside shielding, support structures, or optical components. The hierarchical structure allows for clear organization and reusability of components, while also simplifying modifications to the setup.

In the source configuration stage, users specify the type of radiation source – either point-like or extended (e.g., with vertices distributed in a specific volume) – along with its position relative to the detector and (optionally) the energy of the emitted gamma photons if only a gamma source with fixed energy is simulated. This setup supports both calibration-like conditions and background simulations where radioactivity originates within structural elements. To simulate a specific initial decay kinematics, Simourg uses input data files generated by DECAY0 [6]. This allows users to incorporate realistic multi-particle emissions and energy spectra based on known nuclear decay schemes directly into their simulations. The extended DECAY0 format of input data files allows simulating events with any set of particles and/or nuclei (including radioactive ones) with any initial momenta.

The simulation itself is based on the Geant4 Monte Carlo framework [1]. A user selects a single predefined physics list, which encapsulates all necessary electromagnetic and nuclear processes, among standard Geant4 physics lists.

Upon execution, the simulation produces data on energy deposition within the detector's active volume. These outputs are then used to derive the detector's response function, which reflects its efficiency and spectral response under the defined geometry and source conditions. This streamlined workflow makes Simourg a practical tool for rapid and accurate nuclear detector modeling, suitable for applications in nuclear physics, radiation monitoring, and low-background experiments.

### 3. Using the application

The installation procedure is described in a git repository [7]. The primary installation option currently available for Simourg 2.0 is to compile the source code. The user can do it as simply as the Geant4 basic example compilation. Another option (for Windows 7.0 and later) is downloading the pre-compiled executable file and Geant4 DLL libraries. In any case, downloading the data files from the Geant4 repositories [1] is necessary.

Then, the user can run simulations by executing the command "Simourg USER_SETUP_FILE.mac". The simulation log information is written to the standard output stream and can be redirected to a file for further analysis. The user can control the verbosity level of the log.

Text macro files "*.mac" should consist of the user setup geometry, source parameters, run-time, and readout parameters for every simulation. To prepare a macro file, one can use examples that go along with source code, explore the GitHub ReadMe file, or open the Qt interface and look at /user/* commands description.

At the end of program execution, the primary result will always be the spectrum of deposited energy in the detector. Additionally, one can obtain a VRML file containing 3D track images and various verbosity levels of the output stream, specifically the hit collection that includes precise tracking information.

### 4. New features of Simourg 2.0

Simourg is still under maintenance and updating to correspond experience and vision of many users who work with it. The latest updates are included in the release of version 2.0.

Simourg v2.0 introduces several key features aimed at enhancing flexibility, usability, and visualization capabilities for particle-detector simulations. One of the major additions is the detector intersection checker, which leverages built-in Geant4 methods to automatically identify overlapping volumes during the geometry construction. This ensures more robust and error-free modeling.

A significant upgrade is the readout system, which now allows tracking energy deposition in any user-defined volume, not only in the "Detector" component. This enables broader use cases, such as background studies or material-specific energy loss tracking. The results can be exported in both ROOT and ASCII formats for convenient analysis (Fig. 1). ROOT output includes a TTree structure that can be easily explored using ROOT's TBrowser, allowing users to quickly inspect deposited energy distributions and event-level data (Fig. 2). ASCII output can be divided into independent files (Fig. 1) with a preferred separator or in one file.

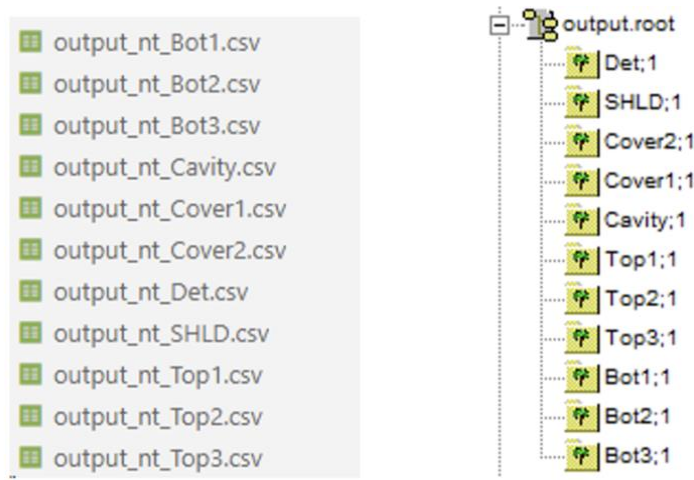

Fig. 1 Possible variants saving the spectra of energy deposition in different volumes.

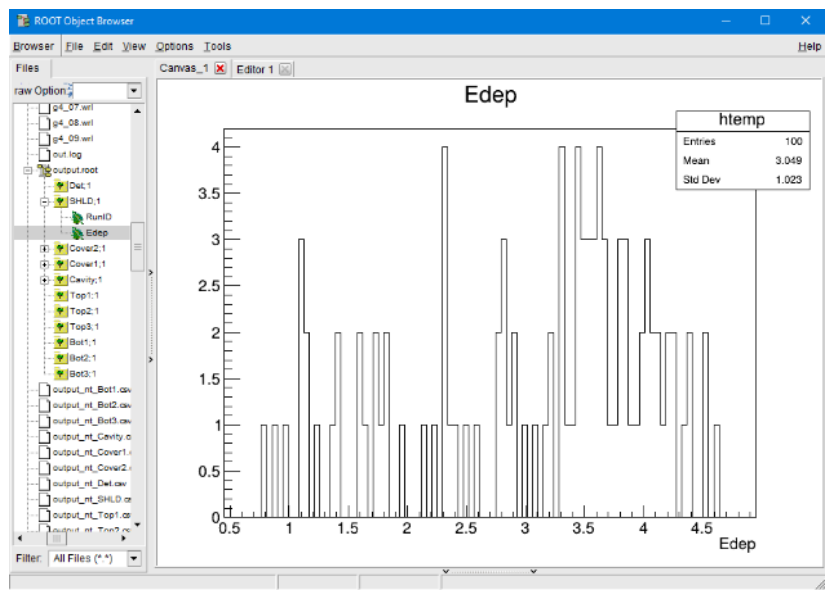

Fig. 2. Visualization of spectra using the CERN ROOT UI.

The precise timing of each energy deposition is captured by the volume hit, enabling the time separation of the detection of daughter particles. Furthermore, the system allows users to write energy deposition per time information for each volume, which can be used to speed up and enhance the analysis process.

On the visualization side, Simourg now offers OpenGL-based geometry rendering with live particle tracking, ideal for debugging geometries and producing clear, publication-ready images (Fig. 3). The VRML output system has also been improved – users can now configure how many VRML snapshots are generated per run, allowing finer control in the debug stage of the simulation.

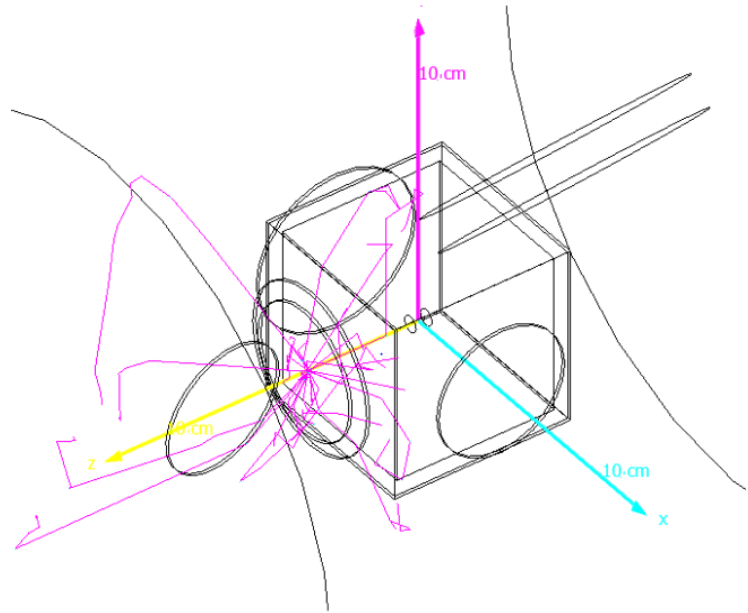

Fig. 3. Project visualization using OpenGL.

Finally, geometry and material definition have become more versatile. Each volume can now be moved in all 3 directions and rotated on 3 axes. The materials of the geometric volume can now be defined through the number of atoms in the chemical formula or by the mass composition of the elements. These improvements enable modeling of more complex experimental setups with off-axis geometries, opening the door to even more realistic and varied detector configurations (Fig. 3).

## 5. Conclusions and usage consideration

Simourg was developed to provide a lightweight, accessible tool for simulating response functions of nuclear detectors using simplified but flexible geometries. By abstracting the complexity of Geant4 into a user-friendly configuration system, it has proven useful in a range of nuclear physics applications, especially where typical geometries and detector layouts are reused or iterated.

With the release of Simourg v2.0, the tool takes a significant step forward. New features such as volume intersection checking, generalized reading of energy releases per time moment in arbitrary volumes, improved output formats (including ROOT and ASCII ntuples), and enhanced visualization options via OpenGL and VRML have expanded its functionality and user experience. The added support for object placement and rotation, and for tuning of material composition, further increases the versatility of geometry construction, keeping the simplicity of the configuration of numerical experiments.

Together, these improvements make Simourg v2.0 a more powerful and flexible platform for modeling particle-spectrometry setups—supporting experimental planning, detector design, and educational applications in a more intuitive and visual way.